\documentclass[10pt,twocolumn,letterpaper]{article}
\usepackage[bottom,multiple,symbol]{footmisc}

\usepackage[pagenumbers]{cvpr} 

\usepackage{bm}
\usepackage{colortbl}
\usepackage{multirow}

\definecolor{cvprblue}{rgb}{0.21,0.49,0.74}
\definecolor{myred}{rgb}{1.0,0.0,0.0}
\usepackage[pagebackref,breaklinks,colorlinks]{hyperref}
\hypersetup{
    linkcolor=myred,
    citecolor=cvprblue,
    urlcolor=cvprblue,
    filecolor=cvprblue,
}

\def\confName{CVPR}
\def\confYear{2025}

\begin{document}
\title{Hybrid Explicit Representation for Ultra-Realistic Head Avatars}

\author{
	{\large Hongrui Cai\textsuperscript{1}\footnotemark[1] \quad Yuting Xiao\textsuperscript{2}\footnotemark[1] \quad Xuan Wang\textsuperscript{3}\footnotemark[2] \quad Jiafei Li\textsuperscript{3}} \\ {\large Yudong Guo\textsuperscript{1} \quad Yanbo Fan\textsuperscript{4} \quad Shenghua Gao\textsuperscript{5} \quad Juyong Zhang\textsuperscript{1}\footnotemark[2]}
	\\ {\normalsize \textsuperscript{1}University of Science and Technology of China \quad \textsuperscript{2}ShanghaiTech University \quad \textsuperscript{3}Xi'an Jiaotong University}
    \\ {\normalsize \textsuperscript{4}Chinese Academy of Sciences \quad \textsuperscript{5}University of Hong Kong}
	\\ {\tt\small hrcai@mail.ustc.edu.cn, xiaoyt@shanghaitech.edu.cn, xwang.cv@gmail.com, jiafeili@stu.xjtu.edu.cn,}
    \\ {\tt\small yudong@ustc.edu.cn, fanyanbo0124@gmail.com, gaosh@hku.hk, juyong@ustc.edu.cn}
}

\twocolumn[{
\renewcommand\twocolumn[1][]{#1}
\maketitle
\vspace{-9pt}
\begin{center}
    \centering
    \captionsetup{type=figure}
    \includegraphics[width=1\linewidth ]{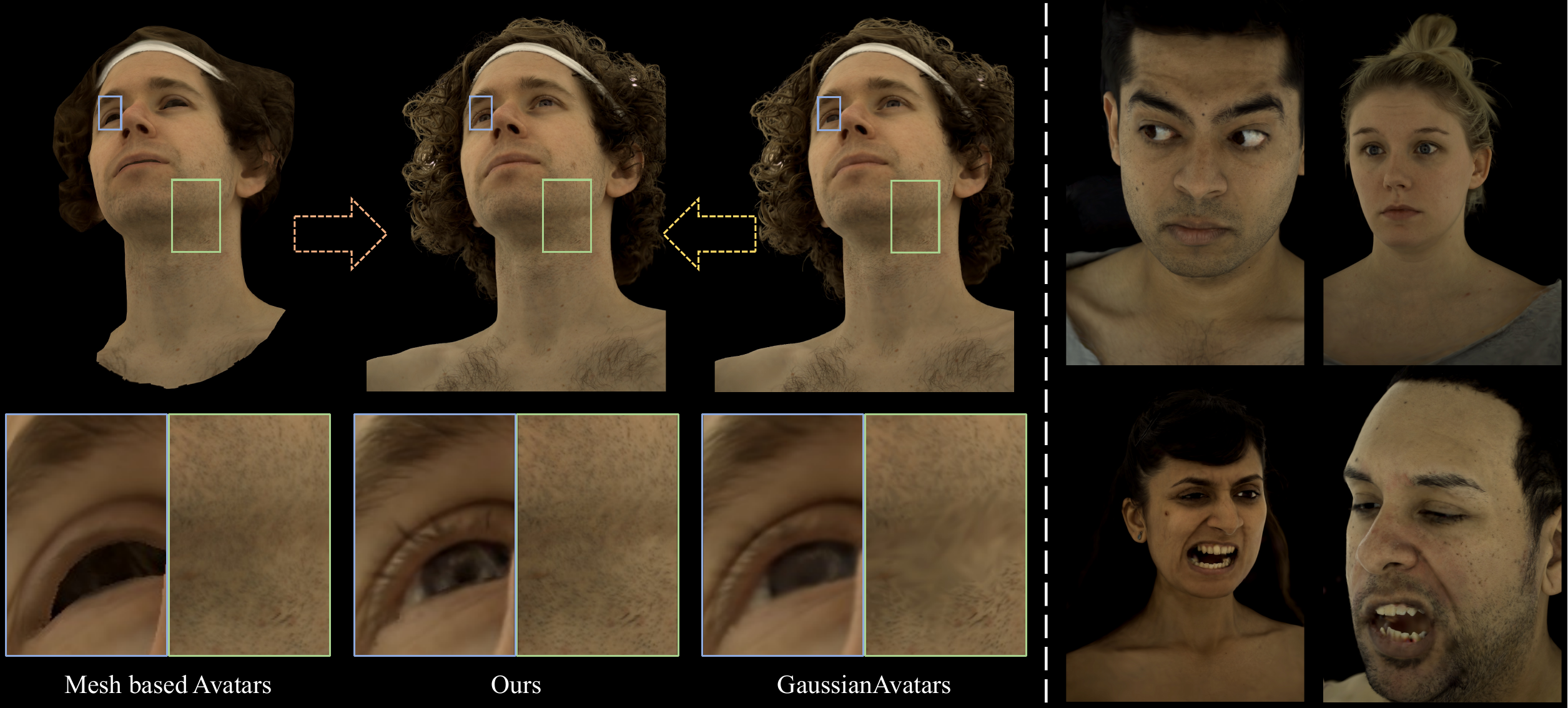}
    \caption{We present HERA, a novel hybrid explicit representation to model 3D head avatars efficiently and effectively. In detail, we combine the advantages of UV-mapped 3D mesh and 3DGS through the proposed hybrid rendering pipeline. On the right side, we showcase the animatable head avatars that employ the hybrid representation, illustrating that high-fidelity rendering effects are achieved across different characters, varying viewpoints and diverse expressions.}
\label{fig:teaser}
\end{center}
}]
\footnotetext[1]{Equal contribution}
\footnotetext[2]{Corresponding authors}


\begin{abstract}
We introduce a novel approach to creating ultra-realistic head avatars and rendering them in real time ($\geq 30$ fps at $2048 \times 1334$ resolution). First, we propose a hybrid explicit representation that combines the advantages of two primitive based efficient rendering techniques. UV-mapped 3D mesh is utilized to capture sharp and rich textures on smooth surfaces, while 3D Gaussian Splatting is employed to represent complex geometric structures. In the pipeline of modeling an avatar, after tracking parametric models based on captured multi-view RGB videos, our goal is to simultaneously optimize the texture and opacity map of mesh, as well as a set of 3D Gaussian splats localized and rigged onto the mesh facets. Specifically, we perform $\alpha$-blending on the color and opacity values based on the merged and re-ordered z-buffer from the rasterization results of mesh and 3DGS. This process involves the mesh and 3DGS adaptively fitting the captured visual information to outline a high-fidelity digital avatar. To avoid artifacts caused by Gaussian splats crossing the mesh facets, we design a stable hybrid depth sorting strategy. Experiments illustrate that our modeled results exceed those of state-of-the-art approaches.
\end{abstract}
\section{Introduction}
\label{sec:intro}

Neural radiance field (NeRF) rendering techniques~\cite{mildenhall2021nerf} have achieved significant advancements in the field of novel view synthesis and have been quickly incorporated into the process of modeling animatable head avatars~\cite{gafni2021dynamic,zielonka2023instant,ma2023otavatar}. Vanilla NeRF techniques use implicit neural representations to depict 3D scenes, which require extensive queries of a Multi-Layer Perceptron (MLP), resulting in inefficient training and rendering. To address these limitations, 3D Gaussian Splatting (3DGS)~\cite{3dgs} has been proposed as a primitive based method that represents 3D content as a set of explicit 3D Gaussians. Many concurrent works~\cite{qian2024gaussianavatars,xiang2024flashavatar,xu2024gaussian} have introduced 3DGS into the construction of head avatars.

Exploiting primitive based rendering in head avatar modeling far predates the 3DGS-specific approaches. In the early stages, the polygon mesh was widely adopted to model digital 3D human heads~\cite{blanz1999morphable,cao2013facewarehouse,thies2016face2face}. Even though the mesh has strong stability in novel view synthesis, the representation ability of mesh is quite limited, especially for reconstructing complicated geometry. In contrast, 3DGS exploits multiple overlapping splats coupled with geometric properties and texture to represent the micro-geometry well under adaptive density control. The 3DGS rendering technique tends to utilize a large number of 3D splats to capture detailed appearance on smooth surfaces. Consequently, compared with the UV-mapped mesh representation, it is easier to produce more ribbed or blurry artifacts in regions of low-frequency geometry combined with high-frequency texture details. Furthermore, it leads to a waste of computational resources to some extent. In summary, it is quite difficult to represent 3D content well with a single primitive type.

In this paper, we present HERA, a 3D head avatar modeling method that proposes a novel hybrid 3D representation consisting of two primitive types: polygon meshes and 3DGS, within an efficient differentiable rendering pipeline. It is strategically designed to leverage the strengths of each primitive type: high-resolution textured polygon mesh which excels at modeling sharp and intricate color appearances on smooth surfaces such as stubble and skin spots, whereas 3DGS performs well in representing complex geometric structures, e.g., hair and eyelashes. We propose a hybrid rendering pipeline to fuse these two primitive types adaptively, leading to better rendering performance than using a single representation alone.

For this purpose, we design a hybrid rasterization procedure. Specifically, for each pixel under a given view, we first apply the mesh rasterization technique to obtain the depths, opacities and colors of intersections of the camera ray with mesh. Then, rasterization concerning the pixel in the 3DGS pipeline yields a list of splats, each with relevant properties. We sort the depth values of the intersections with those of the Gaussian splats. However, we find that directly ordering two-type z-buffers within the target pixel usually causes several Gaussian splats to cross the mesh facets, since the depths calculated in mesh rasterization are theoretically accurate while depths of Gaussians are not per pixel. Therefore, we design a stable depth sorting strategy to avoid creating artifacts in novel view synthesis. After the hybrid rasterization, $\alpha$-blending is applied to obtain the final rendering result.

We adopt the rendering pipeline of hybrid primitives in 3D head avatar modeling. Based on the captured multi-view videos of a person, we track a parametric model, including identity, expressions and poses of each moment. Each Gaussian splat is then parameterized into a local coordinate system of a specific mesh facet in the canonical space and moves together with the rigged facet during the animation process. The colors and opacities of the mesh are modeled in the form of learnable UV maps. We apply the proposed hybrid rendering pipeline on the UV-mapped mesh and Gaussian splats to generate the final result. The proposed HERA optimizes all parameters and densifies the splats by fitting captured color cues from different timestamps and perspectives to ensure that the 3D Gaussians are rigged onto the mesh facet correctly. After optimization, the UV-mapped mesh tends to represent regions of low-frequency geometry with high-frequency appearance, while the rigged Gaussian splats model the intricate and detailed structures, as Fig.~\ref{fig:teaser} shown. Our method outperforms the existing state-of-the-art methods in both novel view synthesis and novel expression animation experiments. Our main technical contributions are as follows:

\begin{itemize}
\item We present a hybrid explicit representation to combine the strengths of different geometric primitives, which adaptively models rich texture on smooth surfaces as well as complex geometric structures simultaneously.
\item To avoid artifacts created by facet-crossing Gaussian splats, we design a stable depth sorting strategy based on the rasterization results of the mesh and 3DGS.
\item We incorporate the proposed hybrid explicit representation into modeling 3D head avatars, which render more fidelity images in real time.
\end{itemize}


\section{Related Works}
\label{sec:related}

\subsection{3D Representations for Novel View Synthesis}
Various 3D representations for novel view synthesis have been extensively studied in recent years. The design of a 3D representation aims to achieve high-quality, efficient rendering while maintaining minimal memory overhead. In these geometric forms, NeRF~\cite{mildenhall2021nerf} applies neural implicit functions to model the radiance field, achieving photorealistic results through volumetric rendering. To generate accurate and smooth surfaces, some following works~\cite{wang2021neus,yariv2021volume,yuting2024debsdf} apply Eikonal regularization for learnable signed distance functions. In reconstructing dynamic scenes, many previous methods~\cite{li2021neural,pumarola2021d,cai2022ndr} achieve promising results without prior templates. Furthermore, several 3D rendering approaches rely on voxel grids~\cite{muller2022instant,neuralvolumes,song2024city} to represent the geometry or appearance of an object. Despite their benefits, implicit representation usually suffers from huge computational complexity, leading to significant resource demands. In the explicit representations, the two main geometric primitives are the polygon mesh and the 3D point cloud. In recent years, differentiable rendering techniques~\cite{liu2019soft,nvdiffrec} have been proposed to recover meshes and textures from 2D images, which are integrated into popular software such as PyTorch3D~\cite{pytorch3d} and Nvdiffrast~\cite{nvdiffrast}. Though the 3D mesh is popular in the applications of computer graphics owing to its flexibility to interact with and edit, obtaining an accurate mesh representation is still challenging. In the production process of industrial design, a considerable amount of manpower and resources are required to create high-precision 3D meshes. By contrast, the 3D point cloud is a simpler geometric primitive used for modeling both geometry and appearance but suffers from temporal instability and holes~\cite{grossman1998point,sainz2004point}. A promising approach, 3DGS~\cite{3dgs}, has been proposed to represent 3D scenes using multiple Gaussian splats. This method, optimized using differentiable rendering, excels at modeling elaborate shapes with high-efficiency rendering. However, it requires a large number of 3D splats to capture both geometric and appearance details, leading to artifacts on smooth surfaces, particularly when the texture is sharp and intricate.

\subsection{Implicit Representations for Head Avatars}
Implicit representations in head avatar modeling~\cite{kirschstein2023nersemble,zielonka2023instant,kirschstein2024diffusionavatars,kabadayi2024gan} leverage the powerful capabilities of deep neural networks, which represent the scene as a radiance field or its compressed forms. NeRFace~\cite{gafni2021dynamic} modulates NeRF based on a latent vector encoded from a monocular video. Further innovation is seen in HeadNeRF~\cite{hong2022headnerf}, which introduces a NeRF based parametric head model that incorporates 2D neural rendering to enhance computational efficiency. The IMavatar framework~\cite{zheng2022avatar} innovates by acquiring a morphable 3D head avatar through neural implicit functions and establishing a conversion from observed to canonical space using an iterative root-finding approach. To model the personalized semantic face, NeRFBlendShape~\cite{gao2022reconstructing} defines the bases as several multi-level voxel fields. AvatarMAV~\cite{xu2023avatarmav} is a fast 3D head avatar reconstruction method using motion-aware neural volumes, employing expression-conditioned voxel grids to describe the deformation field. NPVA~\cite{wang2023neural} proposes a novel volumetric representation based on neural points that are dynamically allocated around the surface for an animatable head avatar.

\subsection{Explicit Representations for Head Avatars}
A classical explicit representation for modeling head avatars is based on meshes, which has led to the development of many models widely used in the industry, such as 3DMM~\cite{blanz1999morphable,cao2013facewarehouse} and FLAME~\cite{li2017learning}. Recently, several points based approaches~\cite{zheng2023pointavatar,qian2024gaussianavatars} have demonstrated topological adaptability in creating lifelike human avatars from scanned data.

Advancements in photorealistic avatar creation~\cite{lombardi2018deep,ma2021pixel,cao2022authentic} have been propelled by the adoption of high-fidelity UV maps and differentiable rendering techniques. Face2Face~\cite{thies2016face2face} marks a pivotal move toward the creation of digital avatars, characterized by real-time face tracking and high-quality facial reconstruction. Subsequent improvements in neural network driven image synthesis methods~\cite{chan2019everybody,thies2019deferred} have augmented the manipulability of head avatars, achieving comprehensive effects from lip synthesis to facial expression transfer and head movements~\cite{kim2018deep,suwajanakorn2017synthesizing}.

Later, NHA~\cite{grassal2022neural} improves upon the FLAME model~\cite{li2017learning}, incorporating subdivision techniques and offsets to refine its geometry and integrating a dynamic texture element that responds to expressions via an expression-dependent texture field. PointAvatar~\cite{zheng2023pointavatar} models the human head as a learned deformable point cloud consisting of locations, normals, and albedo. BakedAvatar~\cite{duan2023bakedavatar} proposes a three-stage pipeline that extracts layered meshes and textures and fine-tunes appearance details with differential rasterization.

GaussianAvatars~\cite{qian2024gaussianavatars} and SplattingAvatar~\cite{shao2024splattingavatar} build a local coordinate frame of a triangle to deform 3D Gaussians on the mesh facets. FlashAvatar~\cite{xiang2024flashavatar} embeds a uniform 3D Gaussian field in a parametric face model and learns additional spatial offsets for modeling regions not on the surface. GaussianBlendshapes~\cite{ma20243d} represents the neutral model and expression blendshapes as 3D Gaussians. MeGA~\cite{wang2024mega} applies occlusion-aware blending on a hybrid representation that combines 3DGS and mesh. However, it renders hairs based on 3DGS and the face based on mesh separately, then stitches them together, while in fact, Gaussian splats are still helpful for fitting facial contours.
\section{Method}
\label{sec:method}

\begin{figure*}[htbp]
    \centering
    \includegraphics[width=0.978\textwidth]{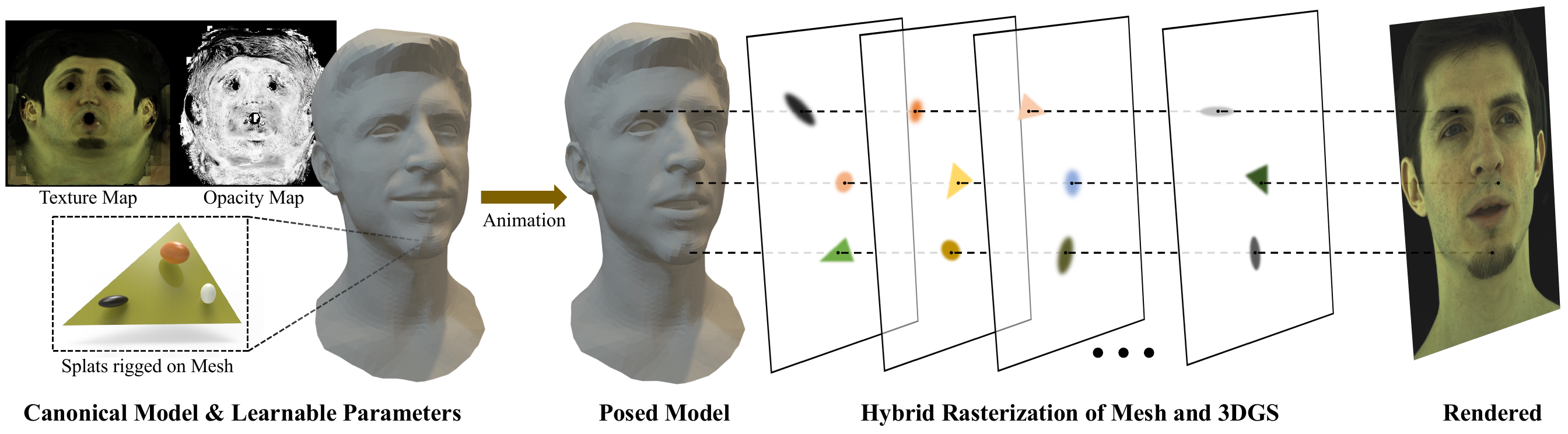}
    \caption{The overall pipeline of proposed HERA. In the canonical space, there is a mesh with a texture UV map $\mathbf{T}$ (visualized in RGB format) and an opacity UV map $\mathbf{A}$, along with several Gaussian splats defined in the local coordinate system of the mesh facets. During animation, the positions of the mesh vertices change, causing the rigged splats to move accordingly. Under the camera view, both the mesh and Gaussian splats are rasterized using the proposed hybrid approach, and the image is rendered through $\alpha$-blending. The entire pipeline is fully differentiable. Guided by the captured image, the texture map $\mathbf{T}$ and the opacity map $\mathbf{A}$ are optimized while the rigged Gaussian splats are updated and densified simultaneously.}
\label{fig:pipeline}
\end{figure*}

The pipeline of HERA is illustrated in Fig.~\ref{fig:pipeline}. First, we briefly review the 3DGS representation. Then, we introduce the hybrid explicit representation, which combines the advantages of 3DGS and polygon meshes, to render scenes with high fidelity and more details. In this section, we present a stable depth sorting strategy for the hybrid rendering pipeline. Subsequently, we rig and parameterize the Gaussian splats into the FLAME mesh~\cite{li2017learning}, enabling human head animation.




\subsection{Preliminary}
Based on calibrated multi-view images as well as a point cloud derived from SfM~\cite{schonberger2016structure}, 3DGS method~\cite{3dgs} optimizes and densifies a set of 3D splats for rendering. Each splat has several properties, including the position (mean) $\bm{\mu} \in \mathbb{R}^3$, covariance matrix $\Sigma$, opacity $\alpha$ and SH coefficients for color $\mathbf{c}$. A Gaussian splat is represented as:
\begin{equation}
    G(\mathbf{x})~= e^{-\frac{1}{2}(\mathbf{x}-\bm{\mu})^{T}\Sigma^{-1}(\mathbf{x}-\bm{\mu})}.
\end{equation}


For radiance field rendering, given a specific viewpoint, the color $\mathbf{C}$ of a pixel is generated by $\alpha$-blending all splats projected onto the pixel in order, as follows:
\begin{equation}
    \label{eq:front2back}
    \mathbf{C} = \sum_{i=1}^{N}
    \mathbf{c}_{gi}\alpha'_{gi}
    \prod_{j=1}^{i-1}(1-\alpha'_{gj}),
\end{equation}
where $\mathbf{c}_{gi}$ is the color of the $i$-nearest Gaussian splat from the camera center, and $N$ is the number of splats overlapping the pixel. The blending weight $\alpha'$ is calculated by evaluating the 2D projection of the splat multiplied by the per-point opacity $\alpha$.

Specifically, during the rendering of Gaussians, the method proposes a fast tile based rasterizer to sort primitives. It is important to note that the rasterized depth value $d_{g}$ of each splat is not calculated per pixel but per tile.

\subsection{Hybrid Explicit Representation and Rendering}
\label{sec:hybrid}
\paragraph{Overview.} Instead of modeling the 3D scene using a single representation, we propose a hybrid representation that combines polygon meshes and 3DGS for mixed rendering. The 3DGS representation excels at fitting complex geometry, while the mesh is ideal for constructing simple and smooth surfaces, especially with subpixel-level visual details. The previous method, SUGAR~\cite{guedon2024sugar}, jointly optimizes the mesh and 3DGS but only utilizes Gaussian splats for rendering. In contrast, our method leverages both 3DGS and mesh representations for hybrid rendering, efficiently modeling the high-frequency appearance of low-frequency geometric structures better than pure 3DGS. Thanks to the very fast rasterization and rendering processes of both 3DGS and meshes, the proposed representation is still able to maintain real-time performance. Additionally, both explicit representations can be naturally integrated into the traditional rendering pipeline.

\paragraph{Hybrid explicit primitives.} The polygon mesh applies the vertices $\mathbf{V}$ and facets $\mathbf{F}$ for geometry as well as UV map $\mathbf{T}$ modeled by SH parameters for appearance. Similar to the 3DGS, which uses opacity values to simulate the radiance field, we also apply an opacity UV map $\mathbf{A}$ to model the $\alpha$-blending process on the mesh, indicating that the facets on the mesh have transparency. Thus, the mesh can be represented as $\mathcal{M} = \{\mathbf{V}, \mathbf{F}, \mathbf{T}, \mathbf{A}\}$.


From a given viewpoint to render a specific pixel, we rasterize the mesh $\mathcal{M}$ to obtain the depth $d_{mi}$, alpha $\alpha_{mi}$, and color values $\textbf{c}_{mi}$ of the $i$-th intersection point on the mesh surface that intersects with the camera ray $\textbf{r}(t)=\textbf{o}+t\textbf{d}$ through the pixel, where $\textbf{o}$ and $\textbf{d}$ represent the camera center and direction, respectively. We record the rendered results of $\mathcal{M}$ as:
\begin{equation}
    \mathcal{R}(\mathcal{M}, \textbf{r}) = \{d_{mi}, \alpha_{mi}, \textbf{c}_{mi}|i=1,2,\cdots, M\}
\end{equation}
where $M$ is the number of intersection points, and $\mathcal{R}$ denotes the mesh rendering process, to obtain the alpha and texture color of each intersection corresponding to the ray.

We sort the overlapping splats of 3DGS and the intersection points on mesh according to the depth values of Gaussian splats $\{d_{gi}|i=1, 2, \cdots, N\}$ and depth values of mesh points $\{d_{mi}|i=1, 2, \cdots, M\}$, where $N$ is the number of overlapping splats of 3DGS defined in Eq.~\ref{eq:front2back}. After hybrid sorting, the alpha and color values of the point list corresponding to each pixel for hybrid rendering are $\{\tilde{\alpha}_k|k=1, 2, \cdots, N+M\}$ and $\{\tilde{\mathbf{c}}_k|k=1,2\cdots, N+M\}$. Based on these results, the rendering function ($\alpha$-blending process) in Eq.~\ref{eq:front2back} is revised to the following form:
\begin{equation}
    \label{eq:hybrid_rendering}
    \mathbf{C} = \sum_{k=1}^{N+M}
    \tilde{\mathbf{c}}_{k}\tilde{\alpha}_{k}
    \prod_{l=1}^{k-1}(1-\tilde{\alpha}_{l}).
\end{equation}

\begin{figure}[htbp]
    \centering
    \includegraphics[width=0.47\textwidth]{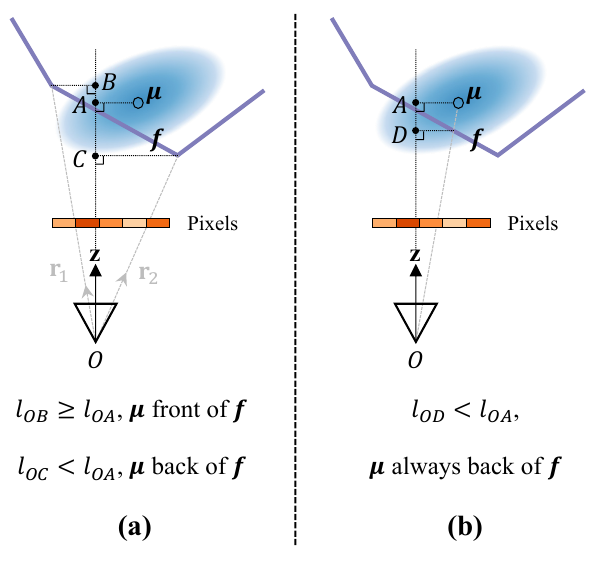} 
    \caption{An example illustrating the stable depth sorting strategy. 3DGS records the depth of a 3D splat (denoted by its mean $\bm{\mu}$) by calculating the distance between its projection onto the $\textbf{z}$-axis and the camera center $O$, \textit{i.e.}, $l_{OA}$, which is not a per-pixel value. During the rasterization of the mesh, individual rays $\textbf{r}_1$ and $\textbf{r}_2$ from $O$ intersect the facet $\textbf{\textit{f}}$ at two points, and the depths of these points are recorded as $l_{OB}$ and $l_{OC}$, respectively. (a) If we directly compare the per-pixel mesh depths $l_{OB}$ and $l_{OC}$ with the splat depth $l_{OA}$, we will place $\bm{\mu}$ in front of $\textbf{\textit{f}}$ when rendering the pixel on $\textbf{r}_1$, but this order is reversed when rendering the pixel on $\textbf{r}_2$. This inconsistency can make the splat appear as if it is crossing the mesh facet, leading to artifacts when rendering from novel viewpoints. (b) Our proposed strategy. Whether sorting along rays $\textbf{r}_1$ and $\textbf{r}_2$, we project the splat onto the mesh facet and compare the depth of the projected point, \textit{i.e.}, $l_{OD}$, with $l_{OA}$.} 
\label{fig:depth_sorting}
\end{figure}

\paragraph{Stable depth sorting.} For the hybrid rasterization of the two types of primitives, we propose a stable depth sorting strategy shown in Fig.~\ref{fig:depth_sorting}. Since the depth of a Gaussian splat is a global value, while the depth of an intersected point on mesh facets is theoretically accurate, directly sorting them at the per-pixel level will cause several splats to cross the facet, creating artifacts under novel viewpoints.

As Fig.~\ref{fig:depth_sorting} shown, to avoid this phenomenon and achieve steady rendering results, we project the mean $\bm{\mu}$ of the 3D splat onto the rasterized mesh depth along the direction of the ray connecting the camera center and obtain the depth of the projected point through bilinear interpolation. We then compare the interpolated value with the depth of the 3D splat to decide whether the splat is rendered in front of or behind the facet. If the projected point is outside the 2D image space, we forcibly set the splat behind the facet in the z-buffer. If the depth of the projected point is $0$, as the ray from the camera center to the splat is not across a mesh facet, we sort them by direct comparison.

Additionally, there is a special case: from a specific viewpoint, the splat needs to be placed behind the mesh facet to represent another part of the scene (such as observing the junction between the strands of hair and the facial contour from the side) but is mistakenly placed in front of the facet according to the designed strategy. To address this, we establish a new rule: if the polygon mesh has a depth value at this pixel and the value is smaller than the depth value of the 3D splat by more than a threshold $\lambda$, the splat is placed behind the facet. We empirically set $\lambda$ to $5$ \textit{cm}. With such a comprehensive design, we are able to achieve stable depth sorting.

\subsection{Head Avatar Animation}
We utilize FLAME~\cite{li2017learning} to model the head avatar parametrically. Following the method that rigs 3D Gaussians onto the FLAME mesh in GaussianAvatars~\cite{qian2024gaussianavatars}, we define a local coordinate system for each facet of the mesh and localize Gaussian splats within it. The location, rotation and scaling of a specific splat are redefined in the local coordinate system. Combined with the geometric attributes of the corresponding facet (e.g., mean edge length, rotation and center), we can obtain the global location of the 3D splat. The splat follows the motion of its rigged mesh facet during head animation. For more technical details, please refer to their paper.




\subsection{Optimization and Implementation Details}
For modeling head avatars, we inherit the basic hyper-parameter settings and the losses from GaussianAvatars, but these are computed on the results via hybrid rendering of two-type explicit primitives. We initialize the texture map $\mathbf{T}$ and the opacity map $\mathbf{A}$ to zero. To accelerate convergence, we first optimize $\mathbf{T}$ and $\mathbf{A}$ in $9,000$ iterations based on rendered results of the pure mesh, then optimize all parameters of 3DGS and mesh representations through the hybrid differentiable rendering. The learning rates of all UV-maps are $5 \times 10^{-4}$. After modeling an avatar based on the proposed hybrid representation, we can interactively drive it and render high-fidelity videos ($2048 \times 1334$ resolution) from free viewpoints in real time, \textit{i.e.}, $\geq 30$ fps, using a single NVIDIA RTX A100 GPU.

\section{Experiments}
\label{sec:exp}

\begin{figure}[htbp]
\centering
\includegraphics[width=0.478\textwidth]{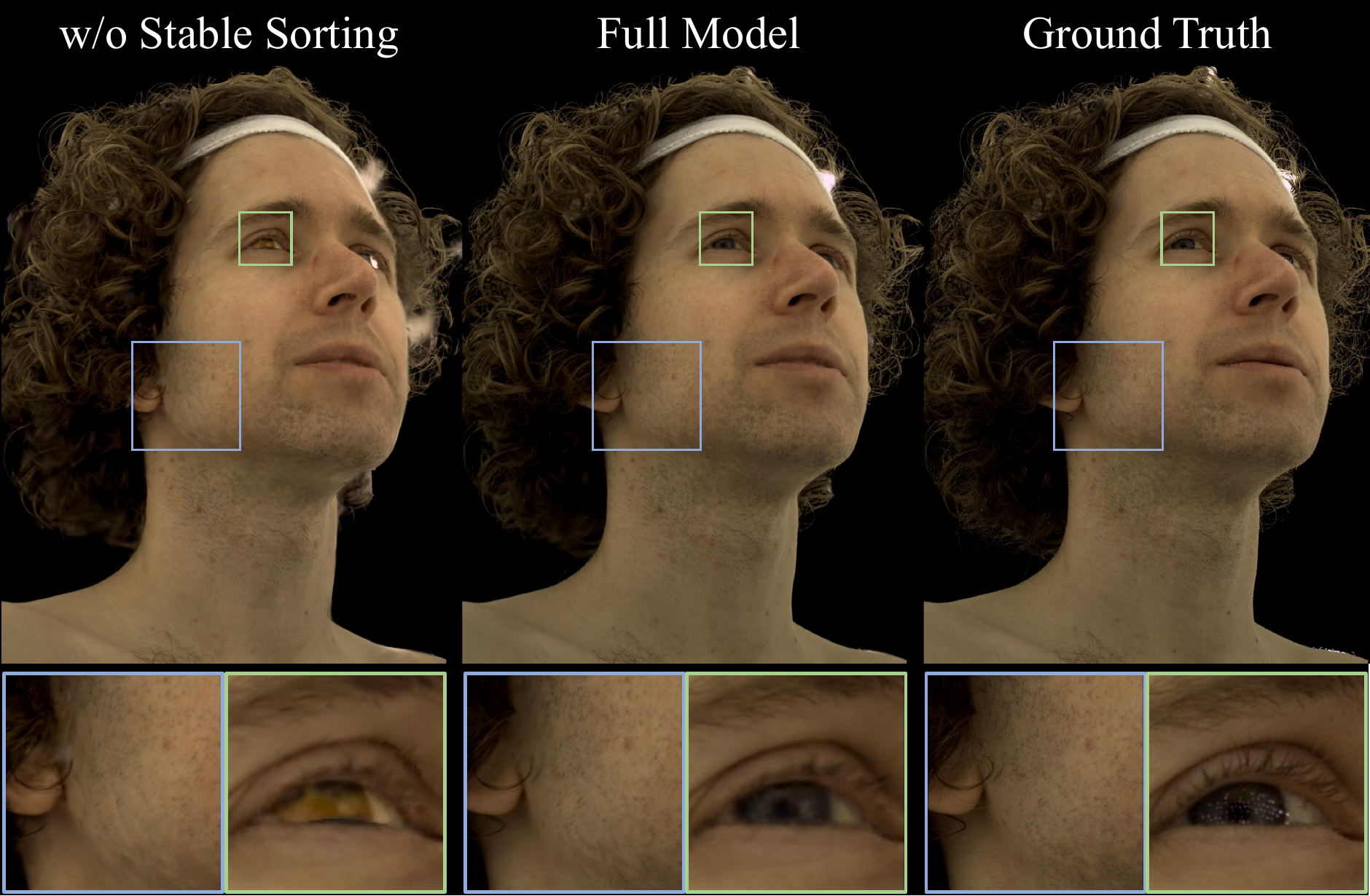}
\caption{Ablation study on depth sorting strategy. It demonstrates that our proposed stable sorting strategy effectively eliminates artifacts in novel view synthesis caused by Gaussian splats crossing the mesh. Zoom in for better views.}
\label{fig:sorting_ablation}
\end{figure}

\subsection{Setup}
\paragraph{Experimental settings.} We evaluate our proposed HERA on $2$ experiments: 1) novel view synthesis by driving an avatar with expressions and poses from training sequences and then rendering from testing views, 2) novel expression animation by driving an avatar with expressions and poses from testing sequences and then rendering from training views. Since parts of the baseline methods require a mask, we compute all metrics between masked results and masked ground truth images for a fair comparison. We utilize the RVM method~\cite{lin2022robust} to obtain the portraits' masks of multi-view videos since it provides a pre-trained model with relatively good perspective and temporal consistency.

\paragraph{Datasets.}
We use multi-view videos at $2048\times 1334$ resolutions of all $13$ subjects from the Multiface dataset~\cite{wuu2022multiface}, including $16$ views for training and $4$ other views for testing. We select a sequence showcasing teeth, a sequence of eye movements, and $3$ speech videos as the training sequences, and chose other $3$ speech videos (not included in the training sequences) as the testing sequences. In the novel expression animation experiments, we choose $4$ out of the $16$ training views.

\paragraph{Baselines.}
For comparisons on the quality of avatar modeling, we select NeRFBlendShape~\cite{gao2022reconstructing}, PointAvatar~\cite{zheng2023pointavatar}, GaussianBlendshapes~\cite{ma20243d} and GaussianAvatars~\cite{qian2024gaussianavatars} as the baseline methods, representing neural implicit functions, point clouds, and 3DGS, respectively. To provide a more comprehensive demonstration, we implement a baseline method based solely on texture-mapped meshes, as Fig.~\ref{fig:teaser} shown. This baseline uses only the tracked FLAME meshes, excluding the eyeballs, with a learnable texture UV map $\mathbf{T}$. No opacity UV map $\mathbf{A}$ is included, assuming the mesh is entirely opaque. We refer to this method as the mesh based avatars.

\begin{table}[htbp]
\resizebox{0.478\textwidth}{!}{
\begin{tabular}{c|ccc|ccc}
\toprule
                   & \multicolumn{3}{c|}{Novel View Synthesis}                                      & \multicolumn{3}{c}{Novel Expression Animation}                                           \\ \cline{2-7} 
\multirow{-2}{*}{} & PSNR$\uparrow$                     & SSIM$\uparrow$                      & LPIPS$\downarrow$                    & PSNR$\uparrow$                      & SSIM$\uparrow$                      & LPIPS$\downarrow$                    \\ \hline
NeRFBlendShape~\cite{gao2022reconstructing}             & 28.841                    & 0.858                   & 0.316                   & 29.358                    & 0.859                   & 0.312                   \\
PointAvatar~\cite{zheng2023pointavatar}        & 29.963                    & 0.864                   & 0.338                   & 27.348                    & 0.847                   & 0.348                   \\
GaussianBlendshapes~\cite{ma20243d}        & 30.288                    & 0.868                   & 0.354                   & 29.332                    & 0.860                   & 0.355                   \\
GaussianAvatars~\cite{qian2024gaussianavatars}     & \cellcolor[HTML]{FFFC9E}31.728 & \cellcolor[HTML]{FFFC9E}0.882 & \cellcolor[HTML]{FFFC9E}0.308 & \cellcolor[HTML]{FFFC9E}34.250 & \cellcolor[HTML]{FFFC9E}0.898 & \cellcolor[HTML]{FFFC9E}0.289 \\ \hline
Ours w/o Stable Sorting              & \cellcolor[HTML]{FFCE93}32.836 & \cellcolor[HTML]{FFCE93}0.898 & \cellcolor[HTML]{FFCE93}0.255    & \cellcolor[HTML]{FFCE93}34.701 & \cellcolor[HTML]{FFCE93}0.906 & \cellcolor[HTML]{FFCE93}0.236 \\
Ours               & \cellcolor[HTML]{FD6864}33.647 & \cellcolor[HTML]{FD6864}0.910 & \cellcolor[HTML]{FD6864}0.232    & \cellcolor[HTML]{FD6864}35.281 & \cellcolor[HTML]{FD6864}0.911 & \cellcolor[HTML]{FD6864}0.221 \\
\bottomrule
\end{tabular}
}
\caption{The quantitative comparisons and ablation study of stable depth sorting on novel view synthesis and novel expression animation on Multiface dataset~\cite{wuu2022multiface}. We denote the \colorbox[HTML]{FD6864}{best}, \colorbox[HTML]{FFCE93}{second best}, and \colorbox[HTML]{FFFC9E}{third best} scores in different colors.}
\label{tab:multiface}
\end{table}

\subsection{Ablation Study: Stable Depth Sorting}
We evaluate the effectiveness of the proposed sorting strategy described in Sec.~\ref{sec:hybrid}. As Fig.~\ref{fig:sorting_ablation} shown, compared to the method of directly comparing the depth values of the mesh and 3DGS (described in Fig.~\ref{fig:depth_sorting}(a)), our stable depth sorting strategy effectively prevents splats from crossing the facets, thus avoiding rendering artifacts in novel views. The quantitative values are presented in the bottom two rows of Tab.~\ref{tab:multiface}.

\begin{figure*}[htbp]
\centering
\includegraphics[width=\textwidth]{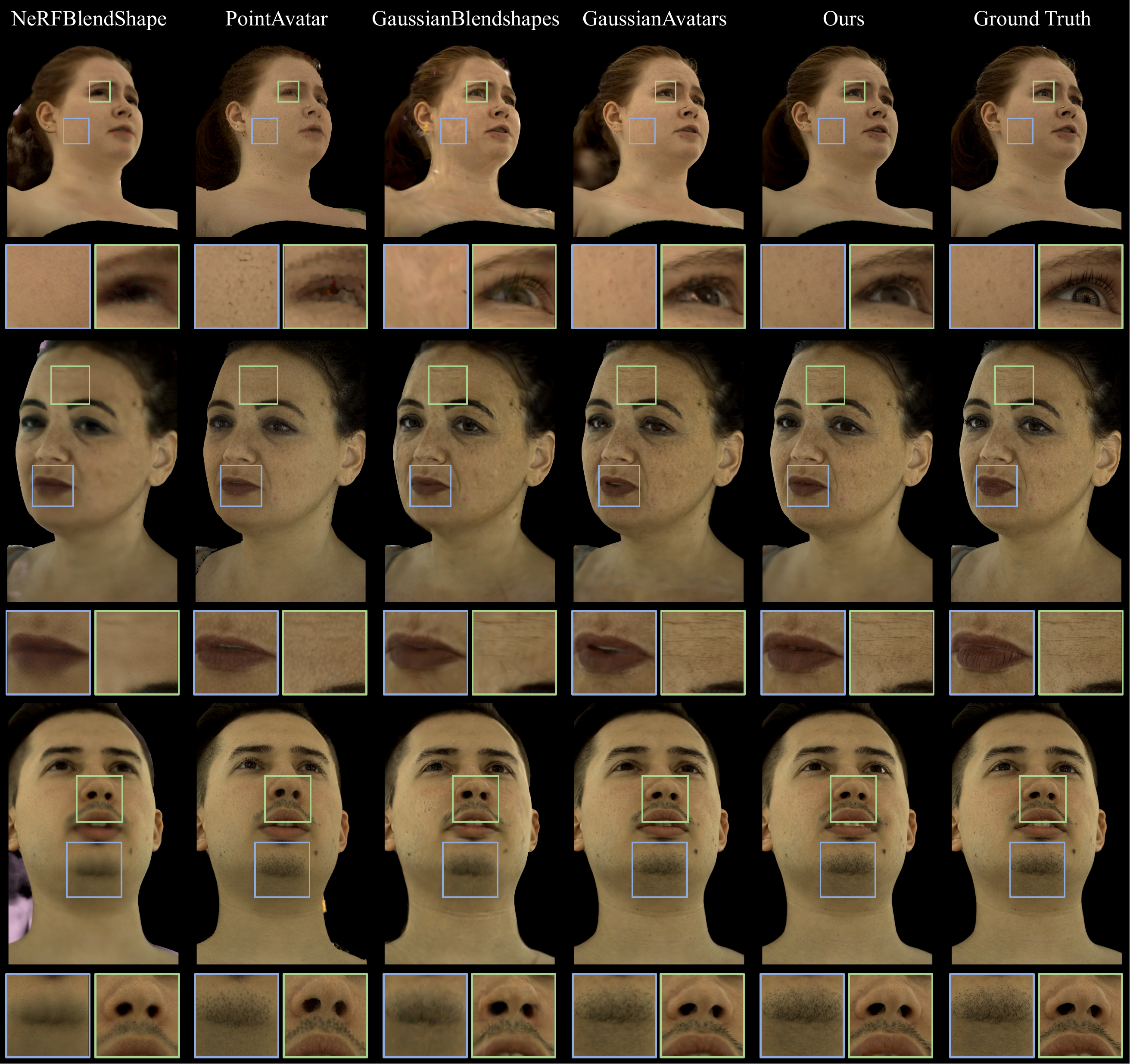}
\caption{Novel view synthesis on Multiface dataset~\cite{wuu2022multiface}. From left to right, we display the results of NeRFBlendShape~\cite{gao2022reconstructing}, PointAvatar~\cite{zheng2023pointavatar}, GaussianBlendshapes~\cite{ma20243d}, GaussianAvatars~\cite{qian2024gaussianavatars}, ours and ground truth images, respectively. Zoom in for better views.}
\label{fig:comparisons_novelview}
\end{figure*}
\begin{figure*}[htbp]
\centering
\includegraphics[width=\textwidth]{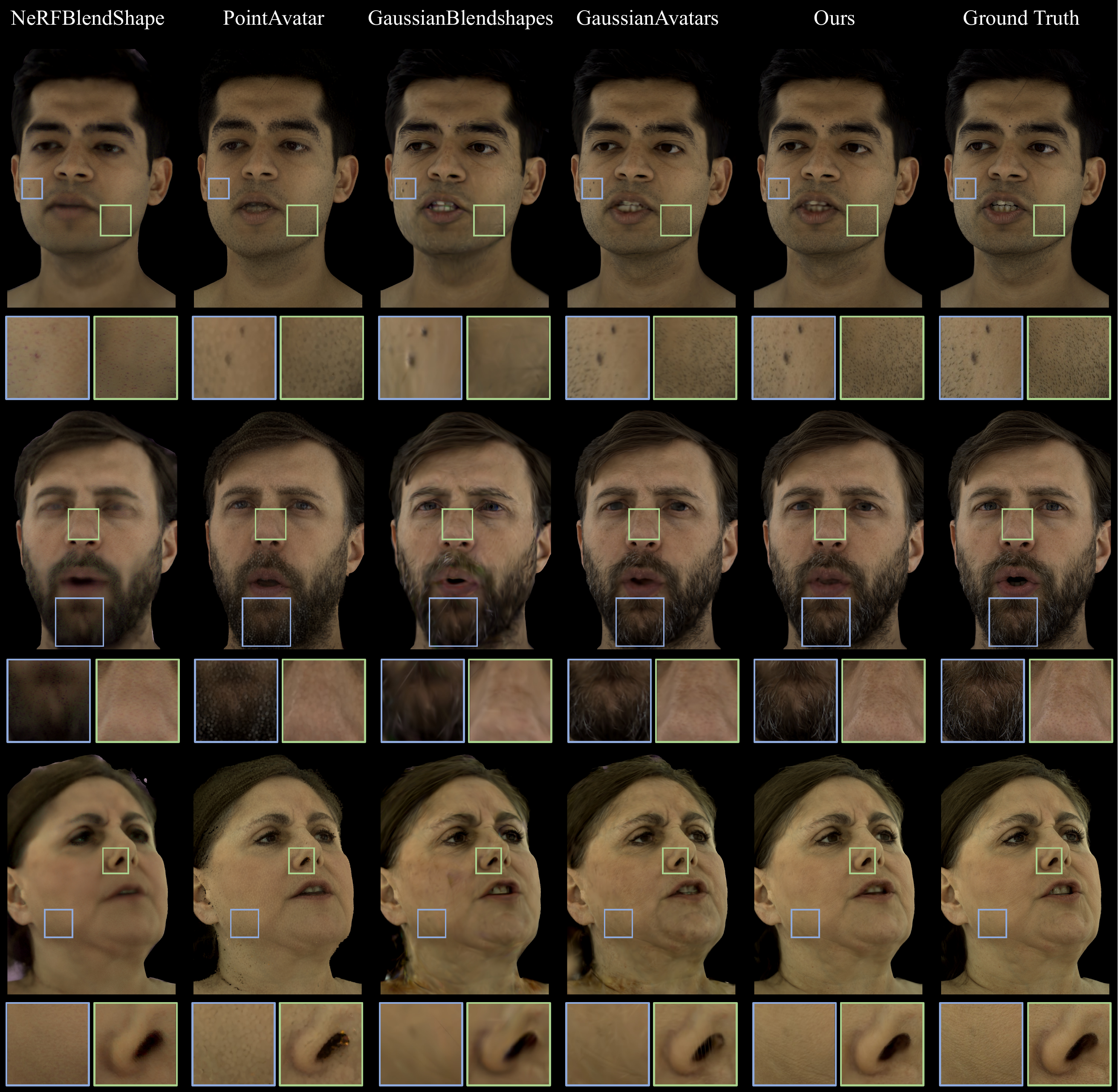}
\caption{Novel expression animation on Multiface dataset~\cite{wuu2022multiface}. From left to right, we display the results of NeRFBlendShape~\cite{gao2022reconstructing}, PointAvatar~\cite{zheng2023pointavatar}, GaussianBlendshapes~\cite{ma20243d}, GaussianAvatars~\cite{qian2024gaussianavatars}, ours and ground truth images, respectively. Zoom in for better views.}
\label{fig:comparisons_novelexp}
\end{figure*}

\subsection{Head Avatar Modeling}
We perform comparisons on the Multiface dataset against representative baseline methods. The qualitative evaluations of novel view synthesis and novel expression animation are illustrated in Fig.~\ref{fig:comparisons_novelview} and Fig.~\ref{fig:comparisons_novelexp}, respectively. The quantitative results are shown in Tab.~\ref{tab:multiface}. We conduct an analysis based on the comparisons of the experiments. 

NeRFBlendShape~\cite{gao2022reconstructing} is a neural implicit function based approach which uses several multi-level hash tables to model the expression basis function. As reflected in the results, it easily loses high-frequency details in learned appearances due to inconsistencies in the driving signals of the training data and performs poorly in synthesizing novel expressions due to the limitations of MLP generalization.

PointAvatar~\cite{zheng2023pointavatar}, GaussianBlendshapes~\cite{ma20243d} and GaussianAvatars~\cite{qian2024gaussianavatars} all utilize explicit primitives as the 3D representation. PointAvatar uses MLPs for geometry and deformation, as well as isotropic points for rendering, which limits its geometric representation ability, resulting in dotted artifacts. Furthermore, it needs plenty of computational resources. GaussianBlendshapes constructs the facial blendshape representation on 3DGS. It generates the avatar corresponding to an expression by combining the neutral model and expression blendshapes through the linear blending of Gaussians with the expression coefficients. It tends to overfit on given views and easily creates artifacts on test views. GaussianAvatars localizes Gaussian splats on the mesh facet and animates them through driving the FLAME mesh. However, compared with textured mesh, 3DGS does not perform well in capturing high-frequency details on smooth surfaces, such as stubble and skin spots.

Our HERA presents a more natural view and expression synthesis over GaussianAvatars while achieving better rendering quality on the skin of the face, thanks to its adoption of the proposed hybrid explicit representation.

For a more comprehensive demonstration, we compare our HERA with the mesh based avatars and GaussianAvatars~\cite{qian2024gaussianavatars} across free views, expressions and poses. Please refer to the supplementary material for the exhibition. It is evident that our HERA combines the strengths of both mesh and 3DGS representations while addressing their respective weaknesses. HERA benefits from the high-fidelity rendering of human faces provided by UV-mapped mesh representation while simultaneously leveraging the detailed modeling capabilities for hair and eyelashes offered by 3DGS.
\section{Conclusion}
\label{sec:conclusion}
We proposed a novel approach to modeling and driving ultra-realistic head avatars. For efficient and high-fidelity rendering, we designed a hybrid explicit representation to adaptively capture different surfaces by blending two types of explicit geometric primitives: mesh and 3DGS. In the rasterization process, we innovated a novel strategy for stable hybrid depth sorting to prevent Gaussian splats from crossing mesh facets. Our proposed approach outperforms state-of-the-art methods in the tasks of novel view synthesis and novel expression animation. Additionally, the utilization of textured meshes offers the potential for various subsequent UV space editing functionalities. However, our method directly outlines the avatar without separating material properties from lighting, so relighting the avatar is not currently achievable with our approach.
\setcounter{section}{0}
\renewcommand{\thesection}{\Alph{section}}
\maketitlesupplementary

\begin{figure}[htbp]
\centering
\vspace{5pt}
\includegraphics[width=0.478\textwidth]{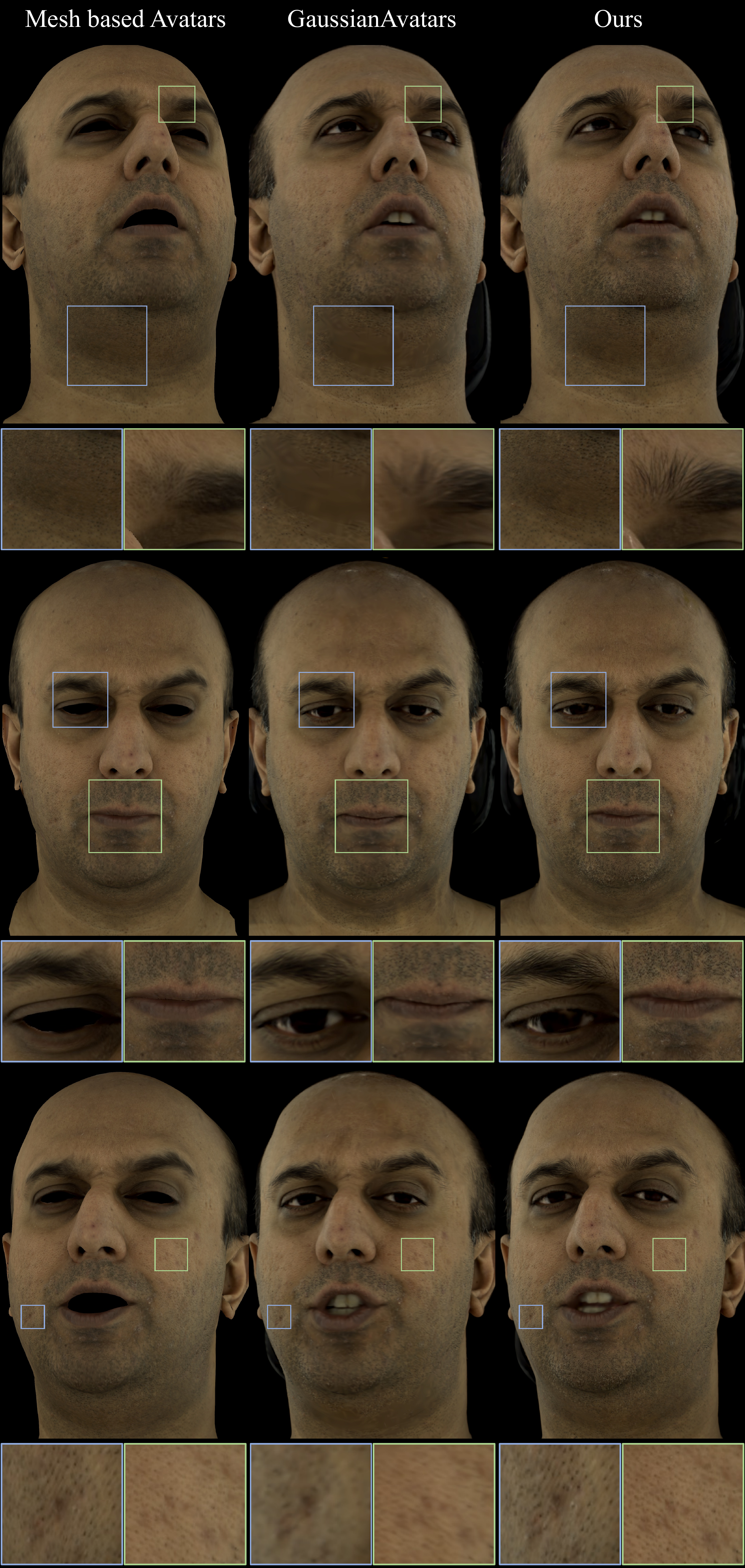}
\vspace{1pt}
\caption{Comparisons of free views, expressions and poses on Multiface dataset~\cite{wuu2022multiface}. From left to right, we display the results of mesh based avatars, GaussianAvatars~\cite{qian2024gaussianavatars} and ours, respectively. Since the viewpoints are freely selected, there are no ground truth images for reference. Zoom in for better views.}
\label{fig:show}
\end{figure}

In the supplementary material, we provide more validations about HERA, including comparisons, parameter quantity and inference efficiency. Then, we introduce two additional tasks to validate our hybrid representation further. For more visual demonstrations, please refer to our video.


\section{More Validations}
As detailed in the Experiments section of the main paper, we implement a baseline that relies exclusively on texture-mapped meshes, called mesh based avatars. We compare our HERA with the baseline and GaussianAvatars~\cite{qian2024gaussianavatars} across various perspectives, expressions, and poses, as illustrated in Fig.~\ref{fig:show}. HERA integrates the advantages of both mesh and 3DGS representations while also overcoming their limitations. HERA takes advantage of the accurate rendering of human faces achieved through UV-mapped mesh, while also utilizing the modeling features for hair and eyelashes available in 3DGS. In contrast, mesh representation struggles to reconstruct complex structural shapes, while 3DGS is less adept at modeling low-frequency geometric surfaces with high-frequency textures.

Furthermore, our HERA utilizes an average of $129,697$ splats to model an avatar from Multiface dataset, which is half the number used by GaussianAvatars (an average of $261,783$ splats). Even if considering the mesh parameters, HERA uses fewer parameters to create a more realistic avatar, which states the proposed hybrid representation makes different primitives reconstruct the scene more efficiently. To infer a video at $2048 \times 1334$ resolution on a single NVIDIA RTX A100 GPU, HERA renders at $81$ FPS which suffices for real-time applications.

\section{Novel View Synthesis on Static Scenes}
\paragraph{Overview.} To further validate the effectiveness of our proposed hybrid presentation, we conduct an experiment on novel view synthesis in static scenes. For this, we follow the settings outlined in 3DGS~\cite{3dgs}. It is important to note that we do not rig the 3D Gaussians onto the mesh facet in this scenario, which means that the association between the two types of primitives is solely established through the hybrid rendering pipeline during the optimization process.

\paragraph{Datasets.} For evaluating static scene data, we perform experiments on the Tanks and Temples~\cite{knapitsch2017tanks} and Mip-NeRF 360~\cite{barron2022mip} datasets. We utilize $5$ scenes from the Tanks and Temples dataset, along with all scenes from the Mip-NeRF 360 dataset in our experiments.

\begin{figure*}[htbp]
\centering
\includegraphics[width=1\textwidth]{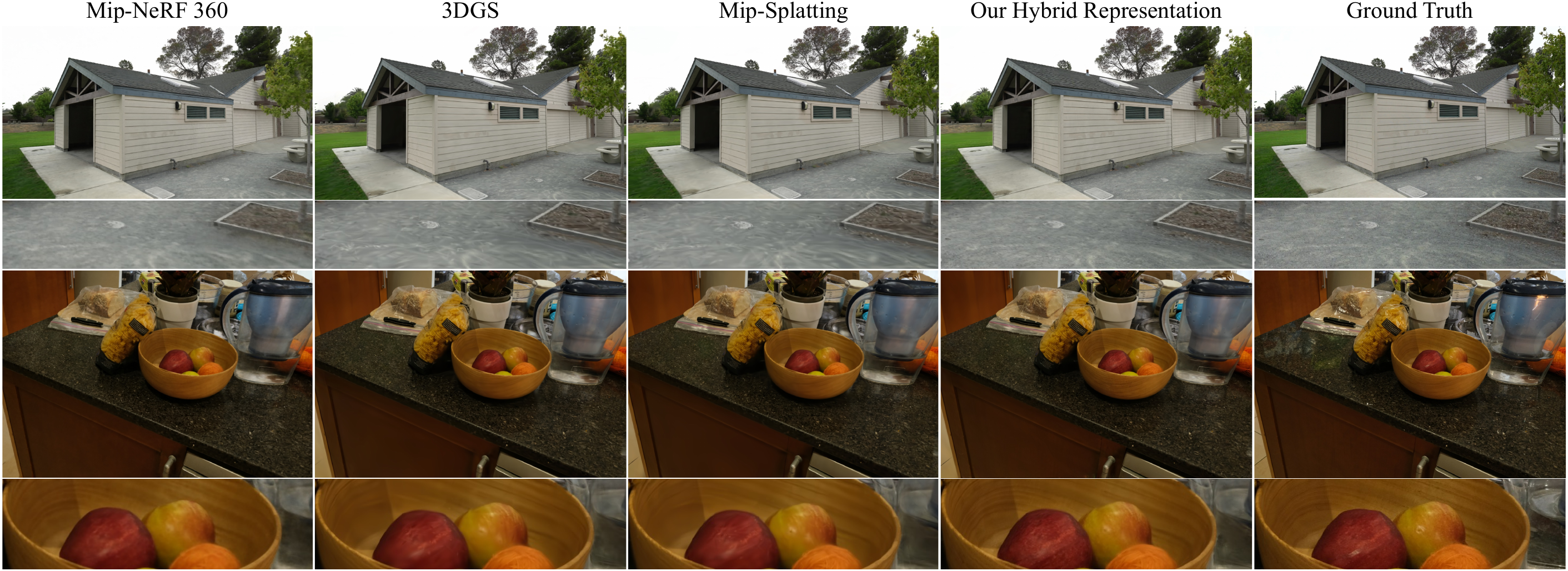}
\caption{Novel view synthesis on Tank and Temples~\cite{knapitsch2017tanks} (top row) and Mip-NeRF 360~\cite{barron2022mip} (bottom row) datasets. From left to right, we display the results of Mip-NeRF 360~\cite{barron2022mip}, 3DGS~\cite{3dgs}, Mip-Splatting~\cite{yu2024mip}, our hybrid representation and ground truth images, respectively. Zoom in for better views.}
\label{fig:static_scenes}
\end{figure*}

\begin{figure*}[htbp]
    \centering
    \includegraphics[width=0.957\textwidth]{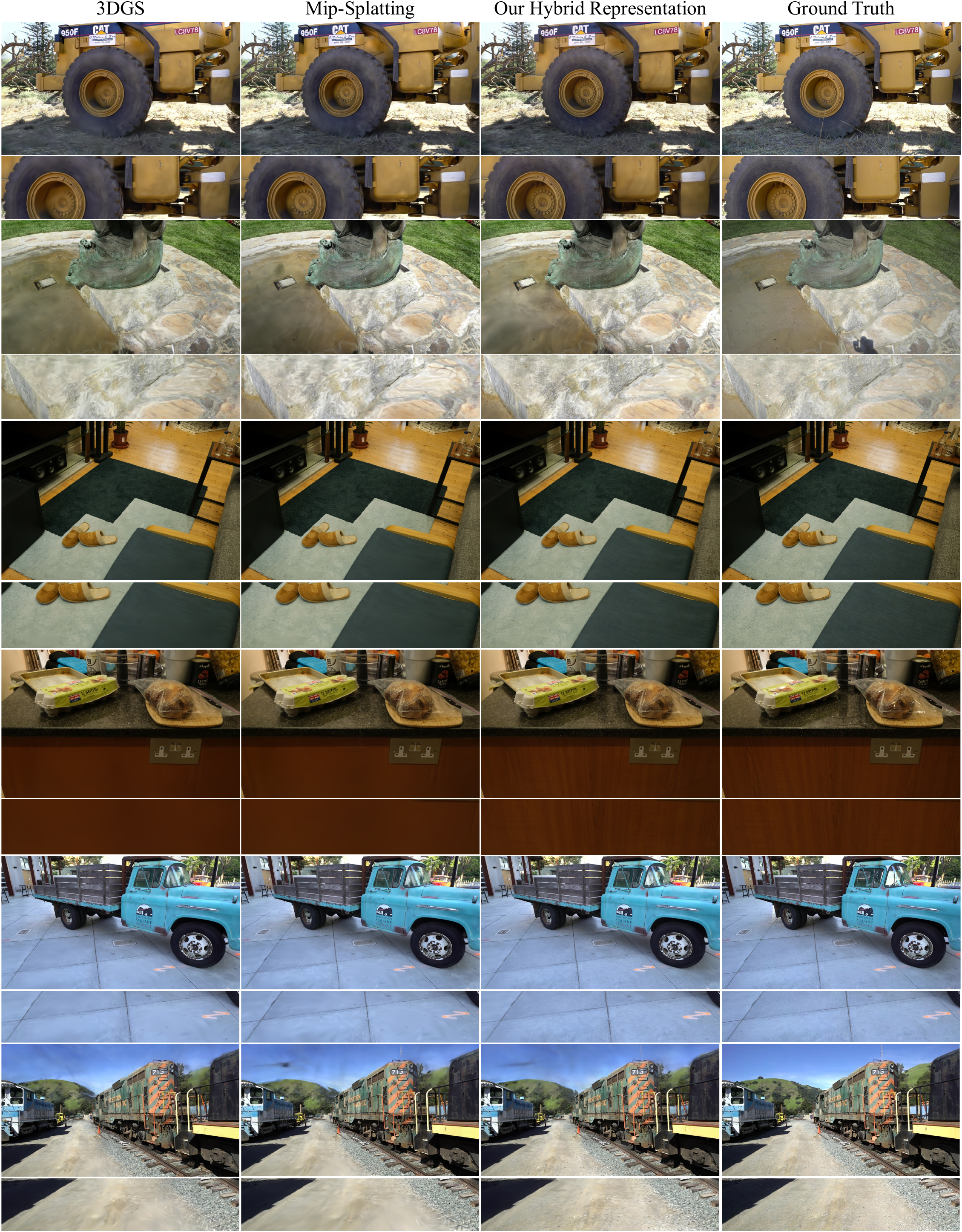}
    \caption{More qualitative results on the static scene datasets~\cite{knapitsch2017tanks,barron2022mip}. From left to right, we display the results of 3DGS~\cite{3dgs}, Mip-Splatting~\cite{yu2024mip}, our hybrid representation and ground truth images, respectively.}
    \label{fig:supp_static_scenes}
\end{figure*}

\begin{table}[htbp]
\centering
\resizebox{0.478\textwidth}{!}{
\begin{tabular}{c|ccc}
\toprule
Method        & PSNR$\uparrow$                         & SSIM$\uparrow$                           & LPIPS$\downarrow$                          \\ \hline
NeRF~\cite{mildenhall2021nerf}          & 23.85                         & 0.605                         & 0.451                         \\
Mip-NeRF~\cite{barron2021mip}      & 24.04                         & 0.616                         & 0.441                         \\
Plenoxels~\cite{plenoxels}     & 23.08                         & 0.626                         & 0.463                         \\
Instant-NGP~\cite{muller2022instant}   & 25.68                         & 0.705                         & 0.302                         \\
Mip-NeRF 360~\cite{barron2022mip}  & 27.57                         & 0.793                         & 0.234                         \\
Zip-NeRF~\cite{barron2023zip}      & \cellcolor[HTML]{FD6864}28.54 & \cellcolor[HTML]{FFCE93}0.828 & \cellcolor[HTML]{FD6864}0.189 \\
3DGS~\cite{3dgs}          & 27.70                         & 0.826                         & 0.202                         \\
Mip-Splatting~\cite{yu2024mip} & \cellcolor[HTML]{FFCE93}27.79 & \cellcolor[HTML]{FFFC9E}0.827 & 0.205                         \\ \hline
Our Hybrid Representation       & 27.61                         & \cellcolor[HTML]{FFCE93}0.828 & \cellcolor[HTML]{FFFC9E}0.199 \\
Our Hybrid Representation*       & \cellcolor[HTML]{FFFC9E}27.75                         & \cellcolor[HTML]{FD6864}0.829 & \cellcolor[HTML]{FFCE93}0.194 \\
\bottomrule
\end{tabular}
}
\caption{The quantitative results on Mip-NeRF 360 dataset~\cite{barron2022mip}. The * indicates incorporating our hybrid representation with Mip-Splatting~\cite{yu2024mip}. We denote the \colorbox[HTML]{FD6864}{best}, \colorbox[HTML]{FFCE93}{second best}, and \colorbox[HTML]{FFFC9E}{third best} scores in different colors.}
\label{tab:static_mipnerf360}
\end{table}

\paragraph{Baselines.} We select Mip-NeRF 360~\cite{barron2022mip}, 3DGS~\cite{3dgs} and Mip-Splatting~\cite{yu2024mip} as the primary comparative baselines. Additionally, NeRF~\cite{mildenhall2021nerf}, Mip-NeRF~\cite{barron2021mip}, Plenoxels~\cite{plenoxels}, Instant-NGP~\cite{muller2022instant} and Zip-NeRF~\cite{barron2023zip} are used for quantitative comparisons.

\begin{table}[htbp]
\centering
\resizebox{0.478\textwidth}{!}{
\begin{tabular}{c|ccc}
\toprule
Method        & PSNR$\uparrow$                          & SSIM$\uparrow$                          & LPIPS$\downarrow$                         \\ \hline
Plenoxels~\cite{plenoxels}     & 21.94                         & 0.708                         & 0.386                         \\
Instant-NGP~\cite{muller2022instant}   & 22.26                         & 0.724                         & 0.347                         \\
Mip-NeRF 360~\cite{barron2022mip}  & 23.61                         & 0.765                         & 0.283                         \\
3DGS~\cite{3dgs}          & 24.01                         & \cellcolor[HTML]{FFFC9E}0.814 & 0.228                         \\
Mip-Splatting~\cite{yu2024mip} & \cellcolor[HTML]{FFFC9E}24.21 & \cellcolor[HTML]{FFCE93}0.822 & \cellcolor[HTML]{FFFC9E}0.216 \\ \hline
Our Hybrid Representation          & \cellcolor[HTML]{FFCE93}24.29 & \cellcolor[HTML]{FFCE93}0.822 & \cellcolor[HTML]{FFCE93}0.207 \\
Our Hybrid Representation*         & \cellcolor[HTML]{FD6864}24.39 & \cellcolor[HTML]{FD6864}0.824 & \cellcolor[HTML]{FD6864}0.206   \\
\bottomrule
\end{tabular}
}
\caption{The quantitative results on Tanks and Temples dataset~\cite{knapitsch2017tanks}. The * indicates incorporating our hybrid representation with Mip-Splatting~\cite{yu2024mip}.} 
\label{tab:static_tanks}
\end{table}

\paragraph{Comparisons.} The quantitative results for Mip-NeRF 360 dataset are presented in Table~\ref{tab:static_mipnerf360}. Our hybrid representation demonstrates significant improvements in LPIPS metrics, achieving comparable performance in PSNR and SSIM metrics relative to state-of-the-art methods. The quantitative results for Tanks and Temples dataset are provided in Table~\ref{tab:static_tanks}. Our representation outperforms the others, showing marked improvements in LPIPS metrics. This can be attributed to the more intricate geometries found in the Mip-NeRF 360 dataset compared to the Tanks and Temples dataset. Many scenes in Mip-NeRF 360 dataset feature extensive grassland areas at the image's center, while the target objects in Tanks and Temples dataset possess smoother surfaces, making them more suitable for mesh modeling.

The qualitative results for static scenes are illustrated in Fig.~\ref{fig:static_scenes}. Our approach shows significant enhancements on smooth surfaces characterized by complex color variations. However, Mip-Splatting still exhibits rendering defects on smooth surfaces, suggesting that these issues are not solely related to anti-aliasing. These experimental results underscore the effectiveness of our proposed hybrid representation. Further qualitative results from the Mip-NeRF 360 and Tanks and Temples datasets are displayed in Fig.~\ref{fig:supp_static_scenes}, demonstrating that improvements are most pronounced on smooth surfaces, which are effectively represented using meshes.

\section{Dynamic Scene Reconstruction}
Our hybrid representation can also model Freeview videos of dynamic scenes. Given a tracked non-parametric mesh, we design a model that utilizes the proposed hybrid representation to jointly optimize the parameters of mesh and 3DGS. As the non-parametric mesh always be incomplete to represent the holistic motion, we apply the tri-plane feature grid to model the deformation field by following 4D-GS~\cite{wu20244d}, instead of rigging splats to the mesh as done in HERA.

\subsection{Methods}
Given the 3D coordinates of a Gaussian, we represent its position, scale, rotation, and opacity in the temporal space by utilizing the deep neural network and tri-plane feature grid~\cite{tensorf,cao2023hexplane,fridovich2023k}. Specifically, given a Gaussian splat in the canonical space, we apply a multi-resolution HexPlane~\cite{cao2023hexplane} and shallow MLP $\phi$ to extract and decode the features from a 4D K-Planes module~\cite{fridovich2023k} which contains $6$ feature planes according to the position in canonical space $\textbf{x}_c=(x, y, z)$ and time $t$. We decode the position, scale, rotation, and opacity of the 3D Gaussian at any given point in the temporal sequence, thereby enabling a robust and detailed representation of the dynamic scene deformation field modeling.

Specifically, the extracted features on the spatial and temporal space are denoted as 
\begin{equation}
\begin{aligned}
        \mathcal{D}(x,y,z,t) = \{& D_l(x,y), D_l(y,z), D_l(x,z), \\
        & D_l(x,t), D_l(y,t), D_l(z,t)| l\in\{1,2\}\},
\end{aligned}
\end{equation}
where $l$ is the upsampling scale and $D_l(i,j)$ is the multi-resolution feature plane. 

After the features $D_l(i,j)$ from multiple feature planes are obtained, they are fused to a global feature vector which consists of both temporal and spatial information:
\begin{equation}
\begin{aligned}
    \mathcal{H} = \bigcup_i\prod \text{interp}(& D_l(y,z), D_l(x,z), \\
    & D_l(x,t), D_l(y,t), D_l(z,t)),
\end{aligned}
\end{equation}
where the "$\text{interp}$" indicates the bilinear interpolation for the 2D feature grid plane. Take the position parameter $\textbf{x}$ of 3D Gaussian as an example, the deformed position can be computed by decoding the feature vector: $\textbf{x}(t)=\textbf{x}_c + \phi_\textbf{x}(\mathcal{H})$. $\phi_\textbf{x}$ is the shallow MLP for decoding the position deformation of Gaussian splat. This design is also applied to the scale, rotation, and opacity parameters of Gaussian splats.

For the mesh representation, since the deformation of mesh has a different pattern from 3DGS, we design the deformation of mesh as the translation of each vertex instead of applying a grid based field, which is a simpler and more direct approach. In this case, each vertex in the mesh has its displacement vector, which directly translates it from its original position to a new position in the deformed state. The mesh based representation contains the vertices, opacity map, and texture feature map for each frame with a shared topological structure. 

In the initial phase of our method, we tackle the challenge of tracking a mesh representation in a dynamic scene using a keyframe based approach. Specifically, we utilize Reality Capture Software to generate a high-quality textured mesh for a selected keyframe. This keyframe mesh serves as the base model against which other frames are registered and aligned. Next, we apply the mesh to a point cloud registration algorithm AMM-NRR~\cite{yao2023fast} to establish the mesh-to-points correspondences and get coarse-tracked meshes to other frames. 

Subsequently, we refine the tracked meshes further in the differentiable rendering framework to finetune the tracked mesh of each frame by applying the photometric loss $\mathcal{L}_{2}$ ensuring that the rendered image closely matches the actual input images from each frame. And the normal consistency loss $\mathcal{L}_{nc}$ to ensure that the surface normals across the mesh faces align well with those estimated from the input data. During this refinement process, the opacity UV map for all the tracked meshes is set to $1$ which indicates full visibility, and the opacity and texture map are not subject to optimization, focusing solely on refining the geometry according to the photometric and normal constraints from the dynamic scene.

The photometric loss for the mesh tracking is defined as the $L_2$ norm term:
\begin{equation}
    \mathcal{L}_{2} = \sum_{\textbf{u}\in\mathcal{U}}||\textbf{C}(\textbf{u}) - \textbf{C}_{gt}(\textbf{u})||^2_2,
\end{equation}
where the $\mathcal{U}$ is the set of the pixels on the image.
The loss function corresponding to this non-rigid tracking phase is:
\begin{equation}
    \mathcal{L}_m = \mathcal{L}_{2} + \lambda_{nc}\mathcal{L}_{nc},
\end{equation}
where the $\mathcal{L}_{nc}$ is the normal consistency loss for smooth regularization to the mesh. 

After the non-rigid tracking of meshes, we initialize the point cloud of 3DGS in canonical space according to the tracked mesh of each frame. For the linear layer $y=Wx+b$ of MLP for decoding the deformation feature vector, the learnable parameters $W$ are initialized as normal distribution $\mathcal{N}(0, \epsilon)$, where the $\epsilon$ is a small value ($1 \times 10^{-4}$), and the learnable parameters $b$ are initialized as zero. This initialization of the deformation field ensures that the deformations are close to zero at the beginning of the training phase, allowing the 3DGS initialized for each frame to be positioned close to the surface. The vertices, opacity map, and texture feature map from mesh representation are all optimized in this phase. 

\begin{figure*}[htbp]
\centering
\includegraphics[width=1\textwidth]{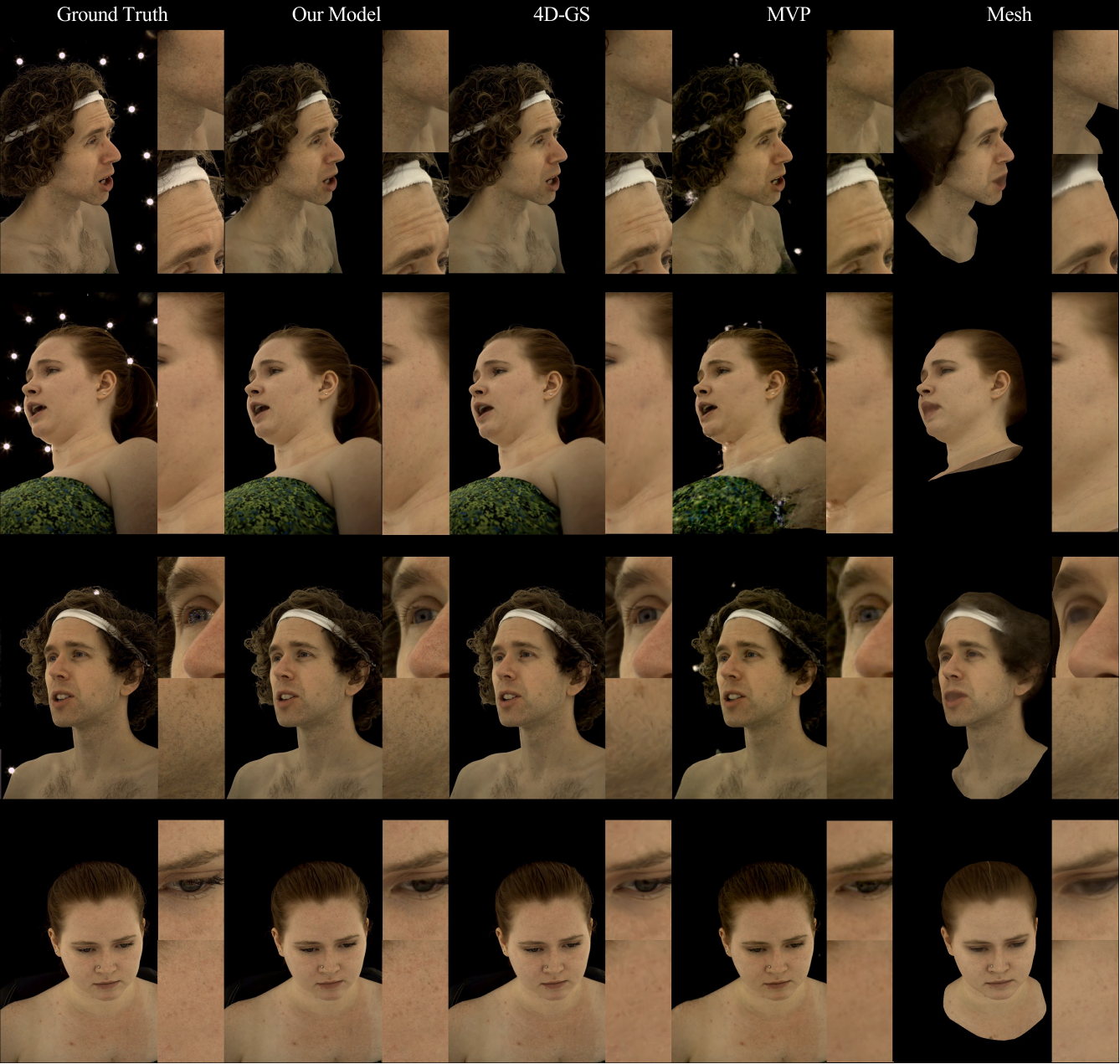}
\caption{The qualitative experiments on the test views of Multiface dataset~\cite{wuu2022multiface} on dynamic scene reconstruction task. From left to right, we display ground truth images, the results of our model, 4D-GS~\cite{wu20244d}, MVP~\cite{lombardi2021mixture} and the tracked mesh, respectively. Zoom in for a better view.}
\label{fig:supp_multiface_freevideo}
\end{figure*}

The loss function of optimizing the hybrid representation is:
\begin{equation}
    \mathcal{L} = \lambda_1\mathcal{L}_1 + \lambda_{SSIM}\mathcal{L}_{SSIM} + \lambda_{tv}\mathcal{L}_{tv},
\end{equation}
where $\mathcal{L}_1$ and $\mathcal{L}_{SSIM}$ is the $L_1$ norm loss and structural similarity loss between the rendered image and ground truth image. The $\mathcal{L}_{tv}$ is the Total Variational (TV) loss on the feature plane of the deformation field on the spatial and temporal range, which follows HexPlane. The loss function is applied to the feature plane of the deformation field across both spatial and temporal dimensions to ensure its smoothness.

\begin{table}[htbp]
\centering
\begin{tabular}{c|ccc}
\toprule
Method     & PSNR$\uparrow$                           & SSIM$\uparrow$                            & LPIPS$\downarrow$                            \\ \hline
Mesh       & 21.56                         & 0.6747                         & 0.2759                         \\
MVP~\cite{lombardi2021mixture}        & 32.98                         & 0.8648                         & \cellcolor[HTML]{FFCE93}0.1612                         \\
4D-GS~\cite{wu20244d} & \cellcolor[HTML]{FFCE93}33.76 & \cellcolor[HTML]{FFCE93}0.8789 & 0.1636 \\  \hline
Our Model       & \cellcolor[HTML]{FD6864}34.23 & \cellcolor[HTML]{FD6864}0.8872 & \cellcolor[HTML]{FD6864}0.1320    \\
\bottomrule
\end{tabular}
\caption{The quantitative comparisons on the Multiface dataset~\cite{wuu2022multiface} for dynamic scene reconstruction. We denote the \colorbox[HTML]{FD6864}{best}, and \colorbox[HTML]{FFCE93}{second best} scores in different colors.}
\label{tab:freeview_comparison}
\end{table}

\subsection{Experiments}
We train and evaluate our model on the Multiface dataset for the dynamic scene reconstruction task. To validate the robustness and effectiveness of our hybrid representation on this task, we have carried out comparative experiments against $2$ baselines: MVP~\cite{lombardi2021mixture} and 4D-GS~\cite{wu20244d}.

The qualitative and quantitative results on the Multiface dataset~\cite{wuu2022multiface} are shown in Fig.~\ref{fig:supp_multiface_freevideo} and Table~\ref{tab:freeview_comparison}, respectively. It can be observed that our method achieves state-of-the-art performance compared with other baselines which indicates its superior performance over existing baselines. This highlights the robustness and effectiveness of our proposed hybrid representation for Freeview video rendering.



Based on modeling the deformation of canonical 3DGS~\cite{3dgs}, 4D-GS~\cite{wu20244d} can effectively reconstruct intricate geometries like hair strands with high fidelity. However, despite its strengths in handling complex structures, this method tends to lose subtle details regarding smooth surfaces, such as the fine whiskers on a human face. 

Meshes are inherently well-suited for representing smooth surfaces since they allow for modeling the surface over vertices and topology, thereby enabling the capture of minute surface variations. 

Our model not only preserves the high-fidelity reconstruction of the hair but also reconstructs the detailed color appearance of the human face. This indicates that our hybrid representation has a significantly better capacity for the details on the smooth surface, which benefits from the mesh representation. The mesh is suitable for representing the smooth surface and the high-resolution texture feature map can achieve excellent modeling for complex color appearance and only occupy very little memory.
{
    \small
    \bibliographystyle{ieeenat_fullname}
    \bibliography{main}
}
\end{document}